\documentclass[preprint]{spie}  

\usepackage{amsmath,amsfonts,amssymb}
\usepackage{graphicx}
\usepackage[colorlinks=true, allcolors=blue]{hyperref}

\title{Experimental challenges for high-mass matter-wave interference with nanoparticles}

\author[a]{Sebastian Pedalino}
\author[a]{Bruno Ramírez Galindo}
\author[a]{Tomas de Sousa}
\author[a]{Yaakov Y. Fein}
\author[a]{Philipp Geyer}
\author[a]{Stefan Gerlich}
\author[a]{Markus Arndt}
\affil[a]{University of Vienna, Faculty of Physics \& Vienna Doctoral School in Physics,  Boltzmanngasse 5, A-1090 Vienna, Austria}

\authorinfo{Send correspondence to:\ sebastian.pedalino@univie.ac.at, \\markus.arndt@univie.ac.at}

\begin{document} 
\maketitle
\begin{abstract}
We discuss recent advances towards  matter-wave interference experiments with free beams of metallic and dielectric nanoparticles. They require a brilliant source, an efficient detection scheme and a coherent method to divide the de Broglie waves  associated with these clusters: We describe an approach based on a magnetron sputtering  source which ejects an intense cluster beam with a wide mass dispersion but a small velocity spread of  $\Delta v/v < 10$\%. The source is \emph{universal} as it can be used with all conducting and many semiconducting or even insulating materials. Here we focus on metals and dielectrics with a low work function of the bulk and thus a low cluster ionization energy. This allows us to realize photoionization gratings as coherent matter-wave beam splitters and also to  realize an efficient ionization detection scheme. These new methods are now combined in an upgraded Talbot-Lau interferometer with three 266 nm depletion gratings. 
We here describe the experimental boundary conditions and how to realize them in the lab. This next generation of near-field interferometers shall allow us to soon push the limits of matter-wave interference to masses up to $10^6$ amu. 
\end{abstract}

\keywords{Matter waves, Quantum mechanics, Matter interference, Quantum optics, Metal clusters, Dielectric nanoparticles, Molecular beams }

\section{INTRODUCTION}
\label{sec:intro}  
Probing  the quantum superposition principle with objects of increasing mass is expected to advance our understanding of the transition from quantum to classical phenomena and thus to tackle the  question why our world appears to be classical.  Matter-wave interferometry is a paradigmatic realization of the superposition principle and both decoherence theory  and a variety of  models of non-standard quantum physics  suggest that deviations from perfect quantum coherence are to be expected for quantum objects of increasing complexity. The influence of novel effects, such as spontaneous or gravitational wave-function collapse\cite{Bassi2003},
conjectured space-time fluctuations\cite{Wang2006} or medium-mass dark matter\cite{Riedel2015,Riedel2017}  are even expected to scale quadratically in mass. Furthermore, certain tests of gravity would profit from the possibility to compare in the same setup the free fall of atoms, clusters and nanoparticles, with different mass, different materials, different internal energy, and many different types of angular momentum\cite{Rodewald2018}.  

\section{INTERFEROMETER CONCEPT}
The overall concept of near-field interferometry goes back to fundamental studies in optics by H.F.Talbot\cite{Talbot1836} who realized that illuminating a single grating by  a plane wave will result in the formation of a grating self-image at  periodic distances behind the mask, as determined by optical near-field interferometry.
For spatially incoherent sources this concept was extended by  E. Lau\cite{Lau1948}  who realized that even for arbitrary sources, sufficient spatial coherence can be created by transmitting  incoherent light through an array of narrow slit sources. This is based on Huygens principle that every point source is the origin of a coherent spherical wave - which is then captured in more mathematical terms by Kirchhoff-Fresnel diffraction and the van Cittert-Zernike theorem\cite{Born1993}.  
Very similar ideas hold for de Broglie waves, where the argument can also be led using Heisenberg's uncertainty relation: enforced confinement to a single slit determined the position which renders the momentum uncertain, and thus enforces quantum uncertainty (coherence) in the  position further downstream.  

\paragraph{Talbot-Lau interferometry:}The combination of  a parallel source array with coherent self-imaging became known as the Talbot-Lau effect and has been used in optics for decades\cite{Patorski1983}. The idea was first implemented into matter-wave interferometry by J. Clauser and S. Li\cite{Clauser1994}  using potassium atoms in an interferometer with three nanomechnanical gratings. Generally, when  three identical transmission masks ($G_1...G_3$) of period $d$ are placed at equal distance, in the vicinity of an integer multiple of the \textit{Talbot length} $L_T=d^2/\lambda_\mathrm{dB}$, the instrument generates a matter-wave density pattern of period $d$ at the position of the third grating, which can be probed by scanning the third mask and counting the number of transmitted particles with an associated de Broglie wavelength $\lambda_\mathrm{dB}$.
Clauser conjectured already in 1997 that this idea might become important in quantum experiments with\textit{ little rocks} and \textit{viruses}\cite{Clauser1997a}. Meanwhile,  research in Vienna  has actually walked  a good part on that path,  in many different experimental realizations.  After the first diffraction of complex molecules\cite{Arndt1999}, the first Talbot-Lau interferometer for molecules was realized with fullerenes\cite{Brezger2002} and porphyrins\cite{Hackermueller2003}. It soon became clear that the velocity-dependent van der Waals interaction between the molecules and material grating walls would limit efforts to upscale this idea to true 'rocks'. This triggered the concept of a Kapitza-Dirac-Talbot-Lau interferometer \cite{Brezger2003}, where the central second grating was realized using a green standing wave of 532 nm laser light retro-reflected at a plane mirror to create a $d = 266$ nm optical phase grating, exploiting the dipole force between the polarizable molecules and the rapidly oscillating electric laser field. The phase imprinted onto the molecular wave function at $G_2$ then results in a molecular diffraction pattern further downstream. This \emph{Kapitza-Dirac-Talbot-Lau} interferometer was realized with larger and floppier molecules  \cite{Gerlich2007} and became also the basis for numerous experiments in molecule metrology, where the interferometer is being used as a quantum sensor for internal molecular properties exposed to external electric, magnetic and optical fields\cite{Gring2010,Eibenberger2011,Fein2020b,Fein2020c}. 

\paragraph{Long-Baseline Universal Matter-wave Interferometer:}
The Long-Baseline Universal Matter-wave Interferometer (LUMI) in Vienna is a three-grating Talbot-Lau interferometer, which started as a 10-fold extension of the Kapitza-Dirac-Talbot-Lau interferometer, with a grating separation of $L=1$ m. This instrument  handled a large variety of different particle classes, from alkali and alkali earth atoms\cite{Fein2020a} over vitamins\cite{Mairhofer2017} and tripeptides\cite{Schaetti2020}, polar and non-polar molecules\cite{Gerlich2007} to molecular radicals\cite{Fein2022}. Recently, LUMI has been used to demonstrate quantum interference of molecules consisting of up to 2000 atoms and with masses up to 28.000\,u~\cite{Fein2019}. All those experiments started from natural or synthetic molecules, naturally predefined nanostructures. And while size and mass-selected nanostructures also occur naturally for instance in the form of proteins or nucleic acids, our universal approach here exploits the in-flight aggregation and post-selection of very massive atomic clusters. A suitable source for these clusters and a useful tool for ionizing them efficiently by ultraviolet light has recently been demonstrated, experimentally\cite{Pedalino2022}: Giant metal clusters of Hafnium and Yttrium, composed of thousands of atoms, can be readily produced in a magnetron sputtering source and photoionized by 266\,nm radiation. LUMI 2.0 now also contains  an upgrade to three ultra-violet photodepletion standing light waves with a period of $133$ nm. This resembles  prior  matter-wave interferometry with polypeptides\cite{Shayeghi2020}, but the use of continuous laser beams in  Kapitza-Dirac diffraction regime adds additional boundary conditions.  Here we focus on how to identify and fulfill the interferometer alignment criteria,  on methods to select the particle velocity using a random photo-depletion chopper as well as on the requirements to normalize signal intensity and phase. 

\subsection{Alignment criteria}

Talbot-Lau interferometry with photodepletion gratings offers a number of important advantages over many other types of matter-wave interferometers: it can accept spatially incoherent particle beams and thus also accept much larger beam divergence angles and orders of magnitude higher molecular flux than far-field interferometry (e.g. Mach-Zehnder interferometry) which would require sub-microradian collimation  of mesoscopic particle beams. Moreover, photodepletion gratings are universal matter wave-front dividers: They are 'threshold' elements rather than resonant matter-optical elements: photons of sufficient energy will always ionize a metal cluster, independent of its internal excitation spectrum. 

Naively one might believe that near-field interferometry with nanoclusters must be terribly hard to align, since the de Broglie wavelength of a 100\,kDa metal cluster travelling at 100\,m/s is only $\lambda_\mathrm{dB}=40$ fm while all foreseeable grating dimensions will always stay above 100 nm, i.e. nearly seven orders of magnitude larger than that. The de Broglie wavelength is also five orders of magnitude smaller than the particle itself and one could believe that even the particle structure should play a role on that scale. Interestingly, it does not -- at least not yet in any of the successful experiments -- because molecule interferometry is always interferometry of the object with itself. However, in order to obtain high interference contrast, strict coherence and alignment criteria have to be met. Some boundary conditions for that have already been described in \cite{Nimmrichter2008,Hornberger2009,Fein2020c}. Here, we adapt them for the specific case of three continuous wave photodepletion gratings acting on slow, massive metal clusters. Moreover, for the following calculations we are considering a beam of yttrium clusters with a mass of $m=100$ kDa and a forward velocity of $v_x= 100$ m/s.

Due to relaxed collimation requirements in Talbot-Lau interferometry compared to far-field diffraction, the goal is to work with as many grating slits $N$ as possible to increase the overall transmitted particle flux. While the use of more grating slits would ideally lead to an increase in the number of Talbot orders and a corresponding increase in the interference contrast, real-world optics impose several boundary conditions that limit the number of Talbot orders that can be achieved. These boundary conditions include factors such as the grating periodicity, spacing, roll, yaw, pitch, and the quality of the diffraction beam profile, as well as limitations imposed by collisional, vibrational, and thermal decoherence. As a result, the number of Talbot orders and the resulting interference contrast may be reduced, and the shape of the interference pattern may be altered. By scanning $G_3$ and counting the transmitted particles, the true particle density pattern can be revealed. If the shape of the interference fringes is sinusoidal, which is typically a valid assumption, the normalized signal difference between the maximum and minimum of the interference fringes can be quantified using the interference fringe visibility:
\begin{equation}
V=\frac{S_{\mathrm{max}}-S_{\mathrm{min}}}{S_{\mathrm{max}}+S_{\mathrm{min}}}. \label{visibility}
\end{equation}This fringe visibility is different (usually higher) for a genuine quantum interference pattern in comparison to a classical Moiré shadow pattern \cite{Hornberger2009}. Understanding the details of all influences is therefore important.

\begin{figure}[h]
    \centering
    \includegraphics[width=0.7\textwidth]{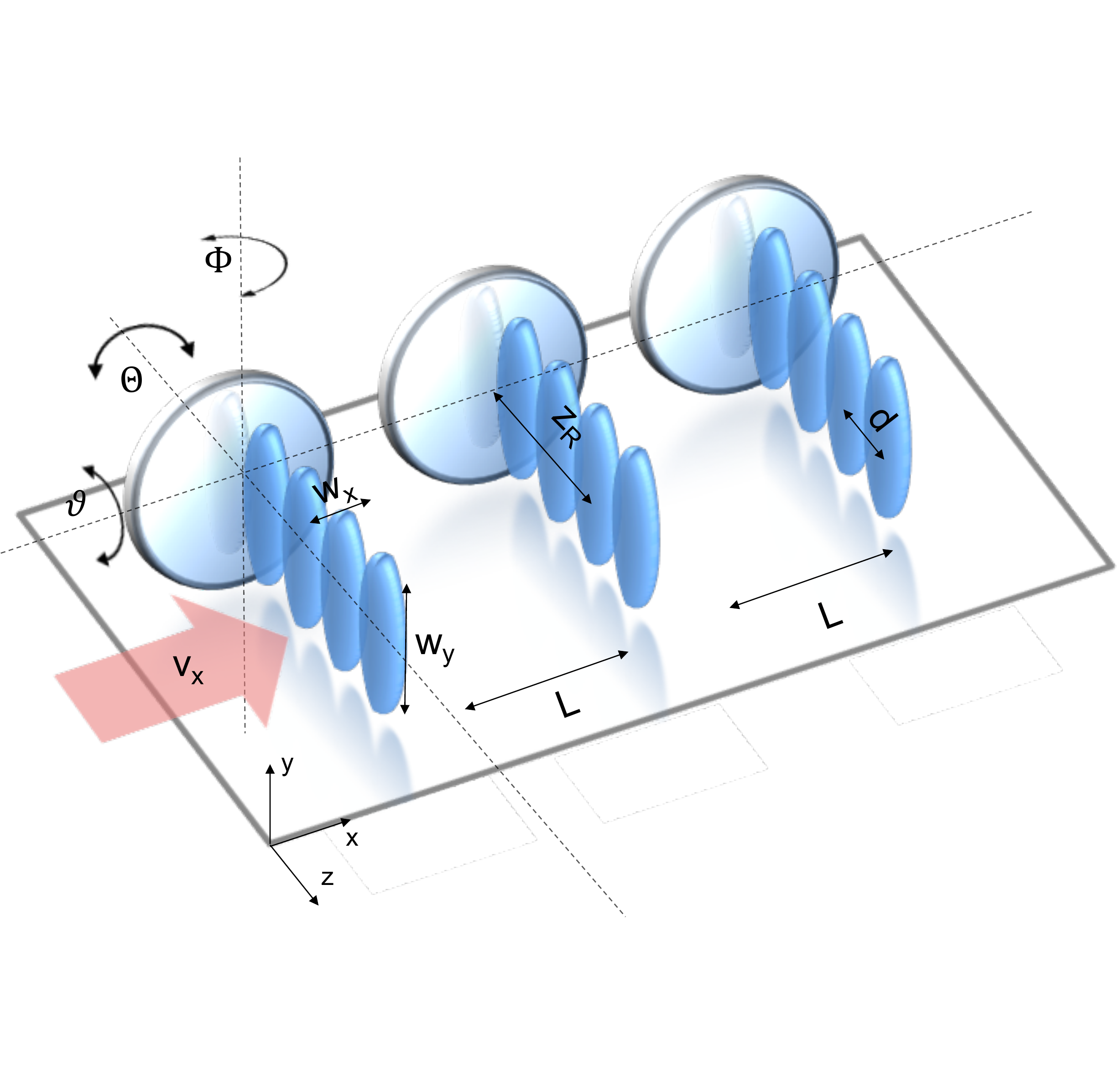}
    \caption{Schematic of the various degrees of freedom for an all-optical Talbot-Lau interferometer. The red arrow indicates the direction of the molecular beam. The grating angles $\vartheta$ (grating roll), $\Theta$ (grating pitch) and $\Phi$ (grating yaw) have to be aligned with respect to the incident molecular beam axis. The vertical grating waist $w_y$ restricts the collimation of the molecular beam height $H$ and the longitudinal waist $w_x$ defines the grating thickness and the ionization efficiency for the clusters with a given absorption cross section $\sigma_{abs}$ and forward velocity $v_x$. The molecular beam should pass the gratings within the Rayleigh length $z_R$,  which is defined by the grating laser wavelength and focal distance of the lens. The grating separation $L$ defines the distance between two optical gratings and is fixed to $L=1$ m. }
    \label{fig:my_label}
\end{figure}

\paragraph{Grating period:} Consider first a wide parallel bundle of geometric (classical) rays passing the three gratings, all separated by the same distance $T_L$.
If all gratings have a perfectly identical period, the matter-wave interferogram will perfectly match the period of $G_3$. However, a  consistent deviation $\Delta d$ from the ideal period $d$  in the first or last gratings will accumulate a fringe shift of $N\cdot d$ when $N$ slits are illuminated. A fringe maximum will thus turn into a minimum for $N\cdot \Delta d=d/2$. A similar argument holds for the genuine interferogram and one is well advised to keep 
\begin{equation}
N < \frac{d}{10\Delta d}  \label{slitnumber}
\end{equation}The laser in LUMI 2.0 has a linewidth of $\Delta \nu =20$ MHz, compared to a base frequency of $\nu=1.127\times 10^{15}$ Hz ($\lambda=$ $266$ nm). This would even allow for  molecular beams as wide as $N=5\times 10^6 d$, that is  gratings larger than 70 cm (!). The high mode quality of modern lasers  is an obvious argument in favor of  optical rather than material gratings, and also an argument in favor of continuous  rather than pulsed laser beams.

 \paragraph{Grating intensity stability:}  Laser powers of approximately 1 W can be consistently achieved at a wavelength of 266 nm, which is sufficient for effectively ionizing and depleting a diverse range of massive metal clusters. However, it has been observed that optical elements tend to degrade over time when exposed to high power UV light, particularly under high-vacuum conditions. While this primarily presents an issue of longevity, mechanical noise and thermal fluctuations within the doubling cavity can also lead to short-term intensity fluctuations, which directly affect the interference contrast.
 
 The interference visibility of the LUMI interferometer asymptotically approaches its maximum with increasing laser power applied at the first and third grating, while it oscillates as a function of power applied at the middle grating.
 In the vicinity of the power optimum, the power dependence of the interference visibility is thus sub-linear for all three gratings. In order to maintain a fidelity of $>90\%$, it is therefore sufficient to maintain: 
 \begin{equation}
      \frac{\Delta I}{I}< 0.1 .
 \end{equation}

\paragraph{Gaussian beam envelope and yaw angle:}
The UV grating lasers are well described by the Gaussian profile with a waist along the molecular beam of $w_x=15$ µm and a waist transverse to it of $w_y=1.5$ mm. This introduces a height dependence in both the laser-induced cluster beam depletion and in the phase shift imprinted in $G_2$. This modulates the fringe visibility but it cannot wash out the interference pattern. Of course,  clusters that do not interact with the diffraction grating only contribute to the background and reduce the fringe visibility. We therefore make sure that the molecular beam height $H$ is smaller than the vertical waist $w_y$, in all three grating zones.

If the cluster beam were highly collimated, the longitudinal waist $w_x$ could also be large, because the accumulated phase is proportional to the laser power P, the optical polarizability  $\alpha_\mathrm{opt}$, the particle forward velocity $v_x$  and the vertical beam waist $w_y$, but independent of $w_x$:  
\begin{equation}
\Delta \varphi \propto \frac{P \cdot \alpha \cdot \tau}{w_x w_y}\propto \frac{P\cdot \alpha}{ v_x w_y}. \label{phase}
\end{equation}A similar argument holds for the absorption and ionization efficiency, which are proportional to the absorption cross section $\sigma_\mathrm{abs}$ rather than to $\alpha_\mathrm{opt}$. And yet a tight focus is needed and  here realized by a cylindrical lens with a focal length of $f=130$ mm to achieve a longitudinal  waist of $w_x=15$ µm.

This constraint is imposed by the finite divergence and also the finite control over the angle of incidence of the cluster beam or the  mirror yaw angle with respect to the standing light wave. The goal is to keep all semi-classical trajectories contained within an angle of incidence $\Delta \vartheta$ that ensures that no cluster can average over a node and a neighboring antinode of the standing light wave:    

\begin{equation}
10\cdot\Delta\Phi < \arctan \left(\frac{d}{4 w_x}\right)\simeq d/4 w_x.
\label{collimation}
\end{equation}
With a period of $133$ nm and a waist of $10$ µm an acceptance angle of $\Delta \vartheta \simeq 0.2$ mrad is still permissible. This allows  a signal enhancement of about $10^6 $  over far-field diffraction at the same grating.

\begin{figure}
    \centering
    \includegraphics[width=0.32\textwidth]{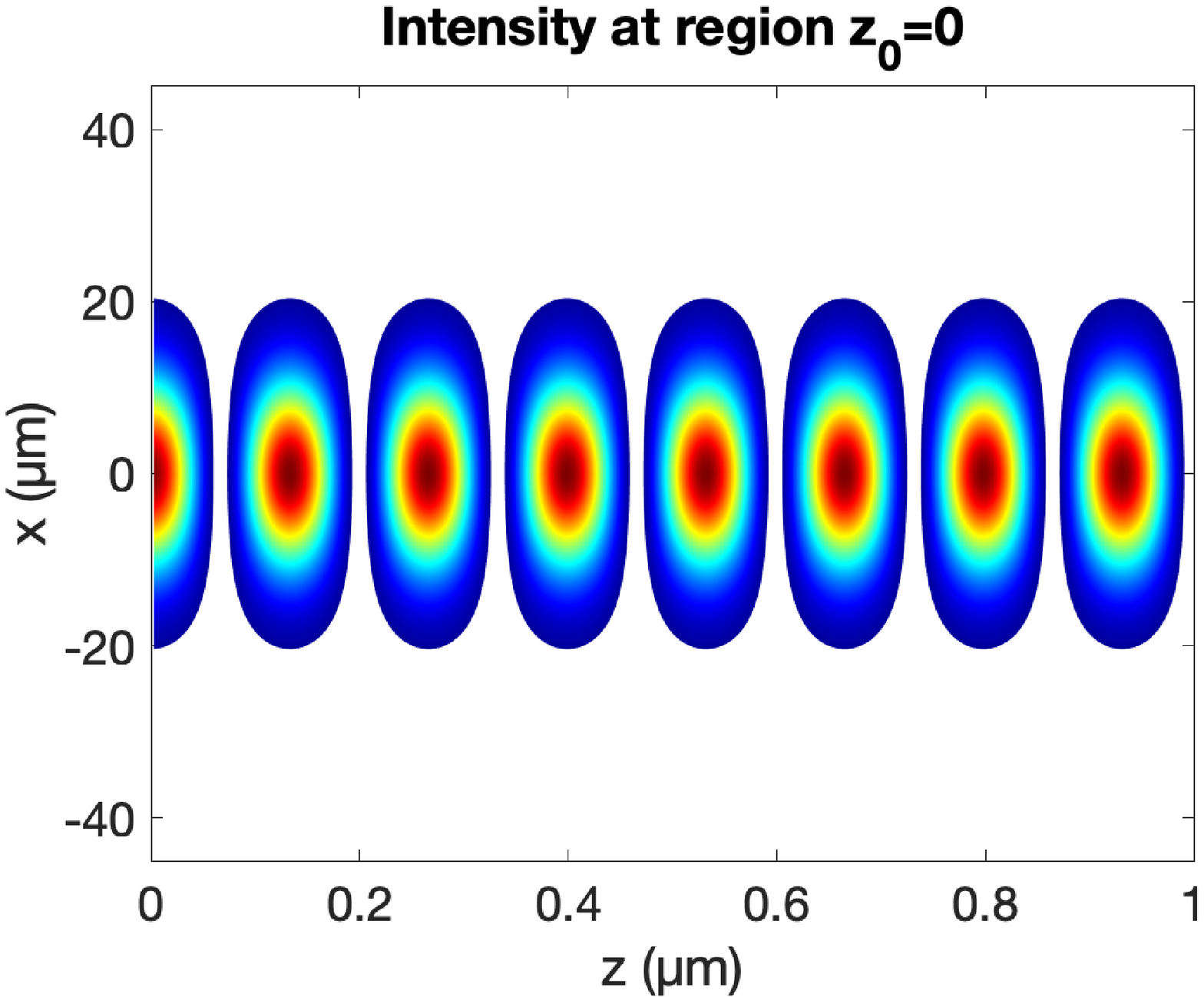}
    \includegraphics[width=0.32\textwidth]{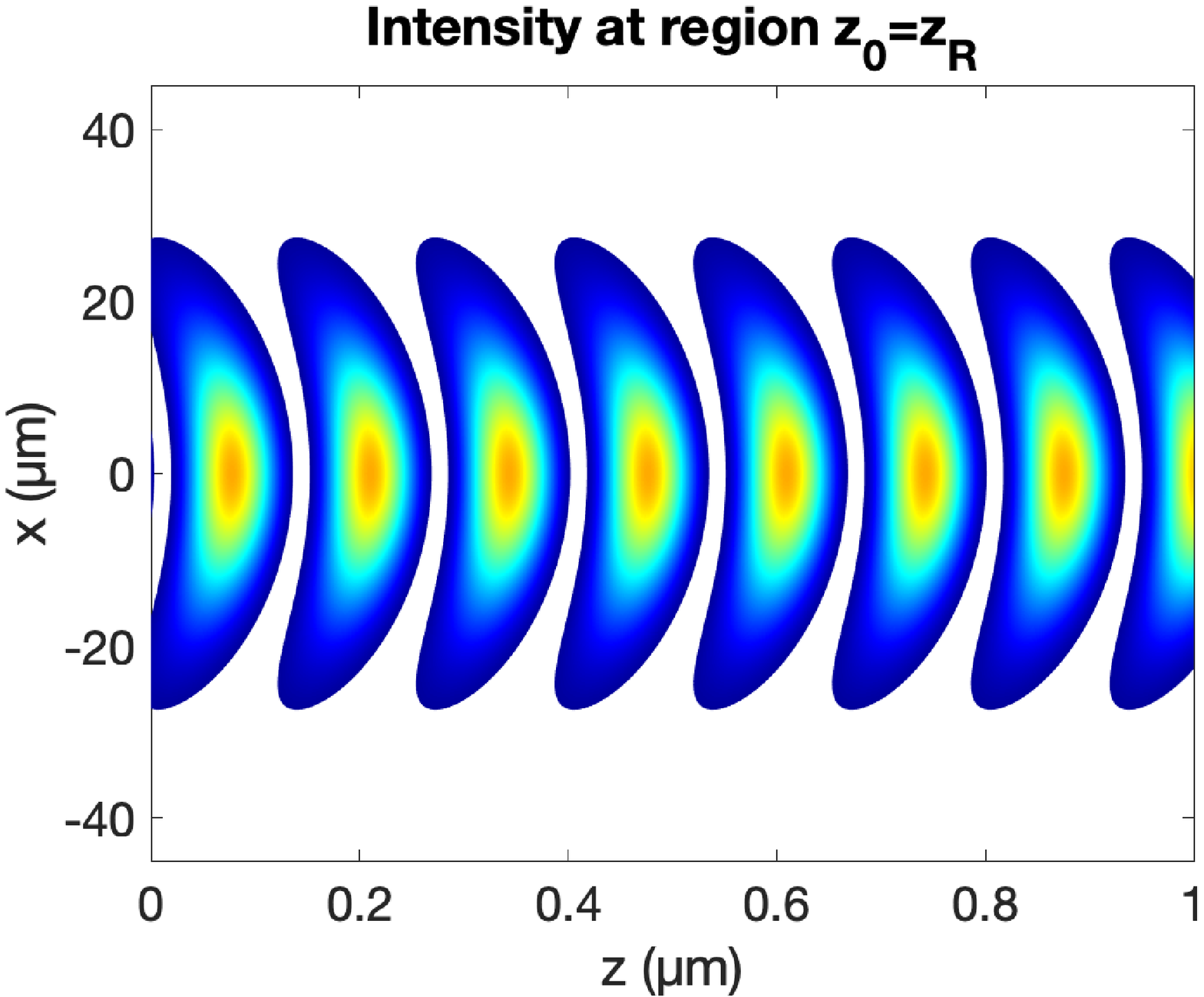}
      \includegraphics[width=0.32\textwidth]{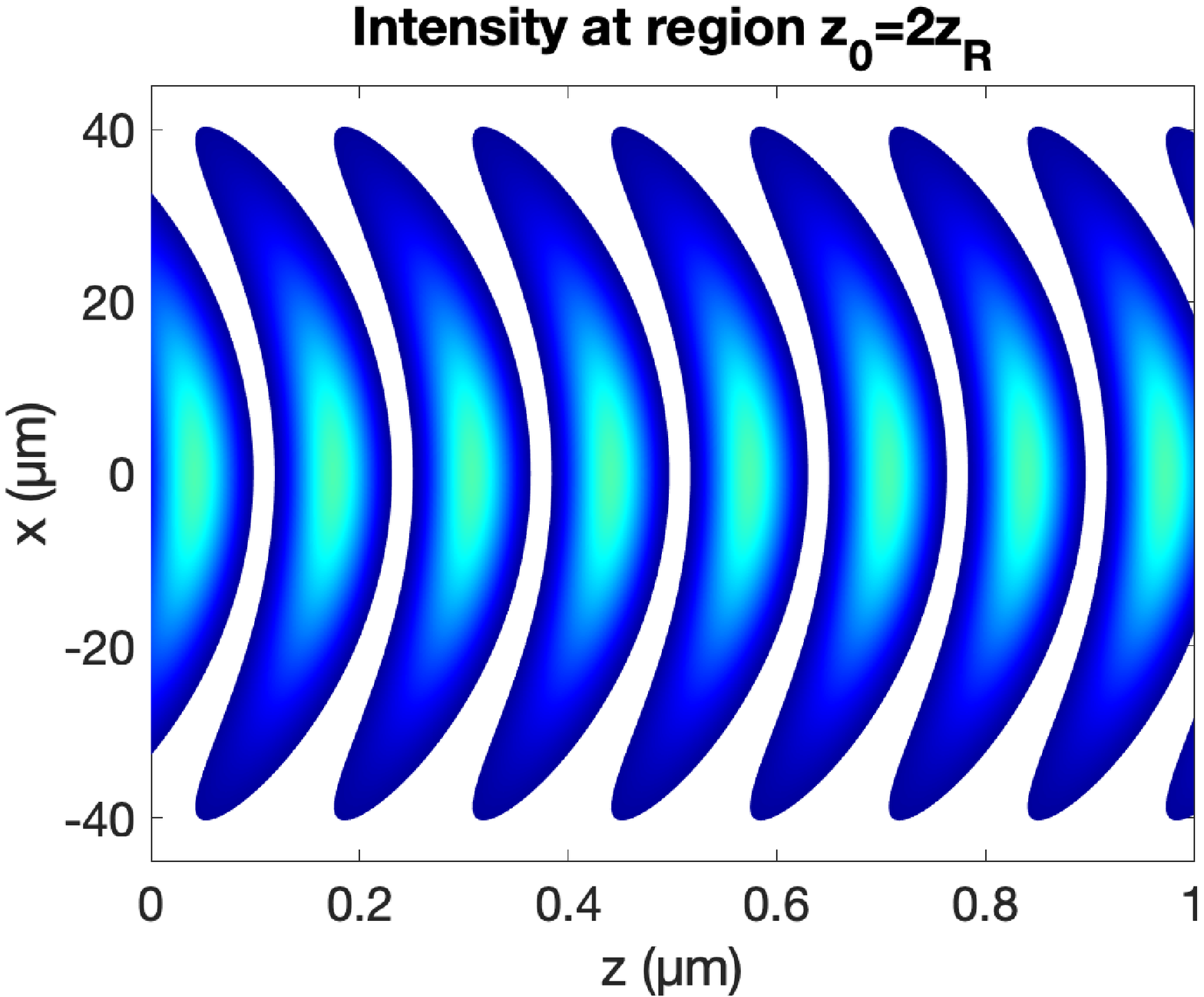}
    \caption{Intensity of the standing light waves at different distances $z_0$ illustrating the effect of the wave front curvature to the grating. As the distance $z$ increases from the focus which is on the mirror surface $z=0$, the radius of the curvature decreases to a minimum at the distance of the Rayleigh length $z=z_R$ and increases again afterwards. However, the waist $w(z)$ is also increasing with $z$ and the resulting wavefronts get more curved the further away one gets from the mirror surface. In order to keep high interference contrast, it is important that the molecules passing through the grating nodes do not see the curved wave fronts of the antinodes. A similar argument holds for the yaw of the grating mirror where the tilt is causing a narrowing of the effective slit width in addition to the curved wave fronts.}
    \label{fig:wavefronts}
\end{figure}

\paragraph{Wave front curvature:}
In Gaussian beam optics, a tight focus $w_x$ is associated with a short Rayleigh length $z_R=\pi w_x^2 /\lambda=2.65$ mm. The beam has a flat wave front on the mirror itself but its radius of curvature changes with distance $z$ from the mirror surface:

\begin{equation}
R(z)=z\left[1+\left(\frac{z_R}{z}\right)^2\right].
\end{equation}
Assuming no beam divergence and a perfect grating yaw we want to make sure that the curvature of the wave fronts does not reach into the nodes of the grating as illustrated by fig. \ref{fig:wavefronts}

\begin{equation}
    w_x(z)\, \tan\!\left[\,\arcsin\!\left(\frac{w_x(z)}{R(z)}\right)\right] < d.
\end{equation}
The molecular beam must pass through the gratings at a distance $z$ that is less than $z_R$, as measured from the point of minimum waist at the mirror surface, in order to fulfill these requirements.

Similar effects can occur due to wave front distortions imprinted by the mirror surface. The laser is focused onto a flat mirror, which alters the wave fronts of the standing light wave. However, these topological changes stay constant with the distance from the mirror and do not spoil the interferogram in any significant matter, as long as the position-dependent shift across the relevant laser beam profile is smaller than $\Delta x < \lambda/10$. This sets a minimum requirement for the mirror surface quality to maintain a clear interference pattern.

\paragraph{Grating separation:}
 Coherent self-imaging in Talbot-Lau interferometry requires precise rephasing of a large set of individual wavelets. 
 For any given velocity, the distance between $G_1$ and $G_2$ as well as $G_2$ and $G_3$ needs to be equal such that\cite{Hornberger2009}

\begin{equation}
\frac{N d}{2L} = \frac{D}{\Delta L} <\frac{d / 2}{\Delta L} \Rightarrow \frac{\Delta L}{L}<\frac{1}{N}.
\label{eq:deltaL}
\end{equation}
For a cluster beam width of $W =1$\,mm, the number of illuminated grating periods is $N=1\mathrm{\,mm}/133\mathrm{\,nm} = 7500 $. We therefore require the relative grating separations to be equal to within $\Delta L < 1/7500 \,\mathrm{m}=133 $ µm.   

\begin{figure}[h]
    \centering
    \includegraphics[width=0.4\textwidth]{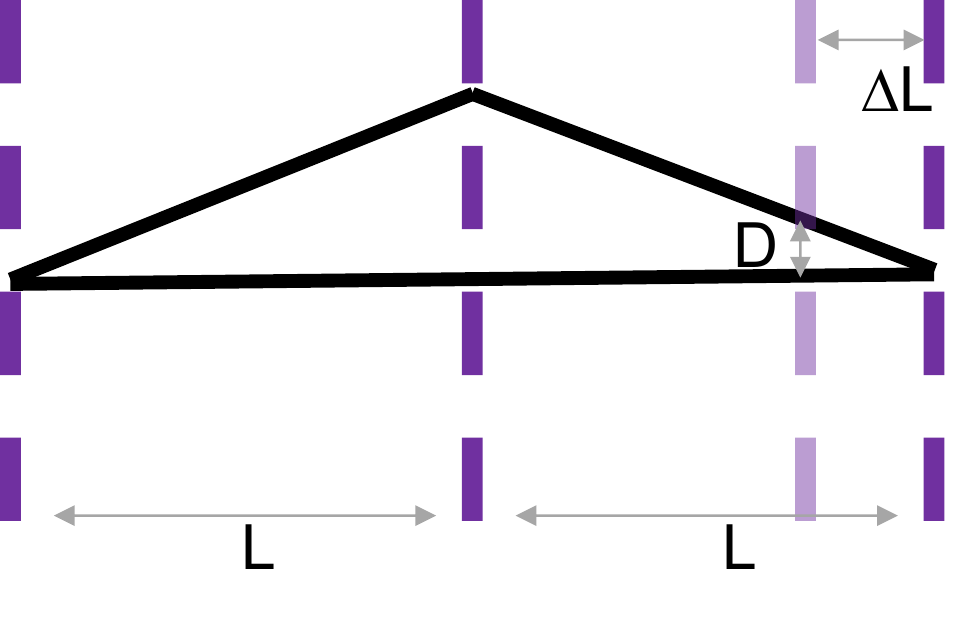}
    \includegraphics[width=0.5\textwidth]{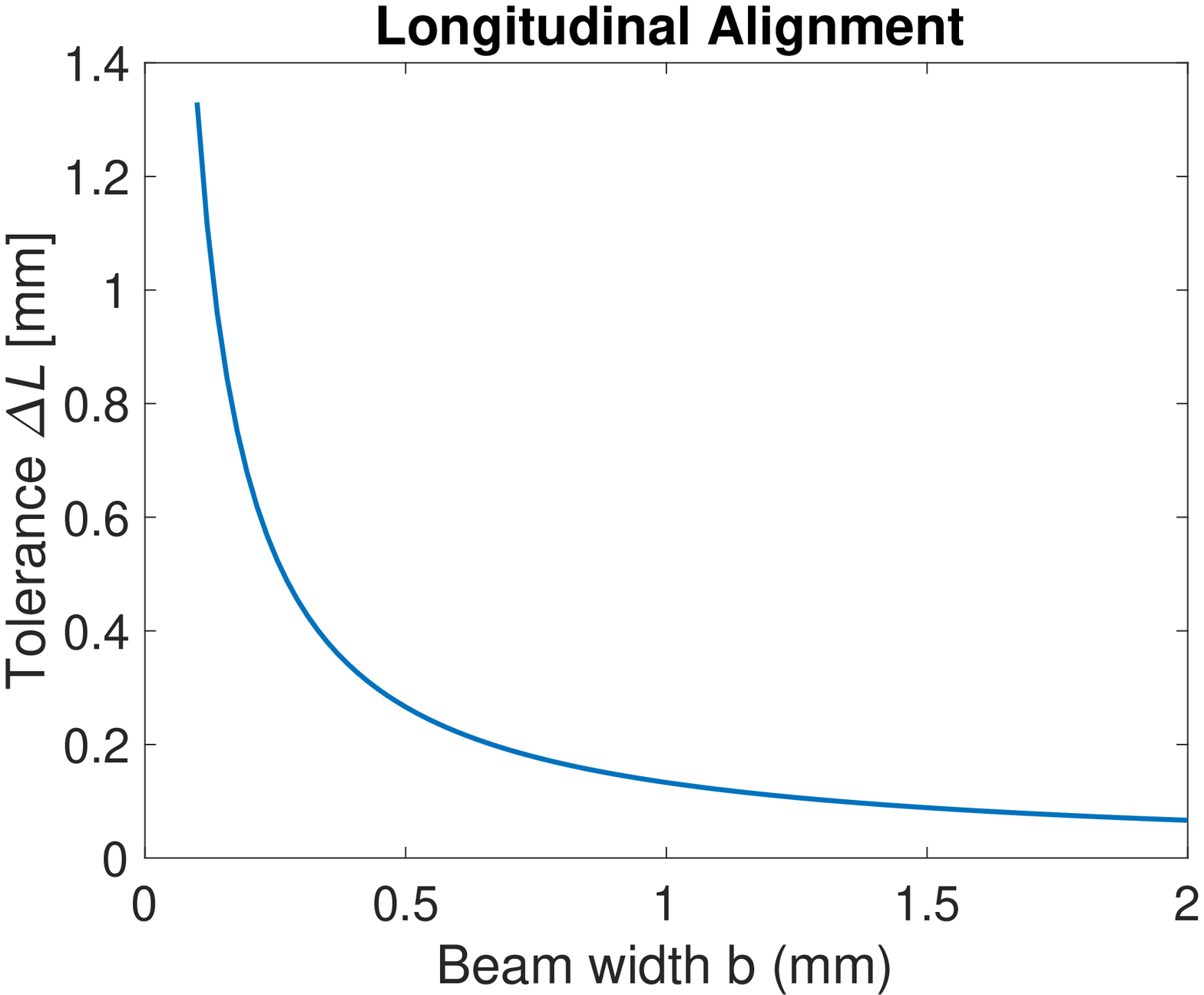}
    \caption{Left: Schematic of a semi-classical trajectory in a three-grating Talbot-Lau interferometer. As the number of illuminated slits $N$ increases, the angle and therefore the diameter of confusion $D$ increases at a given distance. Right: This constrains the grating separations L to be aligned to $\Delta L$ as given by equation \ref{eq:deltaL} for a given molecular beam width $b$ as shown by the plotted graph.}
    \label{fig:my_label}
\end{figure}

\paragraph{Grating pitch:}
All three laser mirrors of our interferometer need to be parallel to each other. Yaw and roll angle are controlled by precise mirror tilts, which can be controlled on the 10 µrad level.  The grating pitch (rotation around the z-axis) is unaffected by any mirror motion, but will be affected by a rotation of the cylindrical lens that focuses the UV beam onto the grating mirror. Aligning this to better than $\theta_\mathrm{pitch}< 100$\,mrad ensures that the ensuing imbalance of the grating distances stays in the  safe regime of 
\begin{equation}
\Delta \theta_\mathrm{pitch} \cdot w_y<  \Delta L     
\end{equation}

\paragraph{Grating roll:}
Grating roll is here defined as a rotation around the x-axis, which points along the cluster beam. In the absence of gravity, the effect of roll is to introduce a y-dependent phase shift. Averaging over this shift reduces the interference contrast by the factor
\begin{equation}
R_{\mathrm{roll}}\simeq\frac{\sin \left(2 \pi \theta_\mathrm{roll} \,H /d\right)}{2 \pi \theta_\mathrm{roll} \,H/d},
\end{equation}
with the cluster beam height $H$, the grating roll angle $\theta_\mathrm{roll}$ and the grating period $d$.

\paragraph{Common roll with respect to gravity:}
In the presence of gravity, particles may fall across different grating slits during their flight time from the source to the detector, even if all gratings are perfectly aligned with each other, if their common roll angle $\theta_g$ with respect to gravity is nonzero. The phase shift is proportional to the effective falling distance $H\propto T^2=L^2/v^2$, which can be huge for slow clusters and a long flight path of $L=1$\,m. However, only the average over different phase shifts reduces the fringe visibility. The reduction of the matter-wave fringe visibility  therefore additionally depends on the relative velocity spread $\sigma_v/v$:
\begin{equation}
R_{\mathrm{grav}}=\exp \left[-8\left(\frac{\pi g \sin (\theta_g) L^2 \sigma_{\mathrm{v}}}{ v^3 d}\right)^2\right].
\end{equation}
This  sets a bound on the maximally permissible roll angle that is compatible with a visibility reduction of less than $R_{\mathrm{grav}}$:  
\begin{equation}
\theta_g<\arcsin \left(\sqrt{-\frac{\ln \left(R_{\text {grav }}\right)}{8}} \frac{d\, v_z^3}{\pi g L^2 \sigma_{\mathrm{v}}}\right).
\end{equation}

\paragraph{Coriolis force:}
A matter-wave interferometer with vertical grating lines, i.e. k-vector perpendicular to gravity, can suppress all gravitationally-induced phase shifts, 
but it will still accumulate a dispersive phase shift because the Earth is rotating underneath the interferometer while the clusters follow their geodesic path in high vacuum:  
\begin{equation}
    \Delta \varphi_\mathrm{Coriolis}=k_d (2 \mathbf{v}\times \mathbf{\Omega}_E) L^2/v^2
\end{equation}
 Here, $k=2\pi /d$ is the grating wave vector, $v$ the forward velocity in the cluster beam and $\Omega_E$ the vector of the Earth's angular frequency. 
Again, an offset velocity does not harm the interference contrast. However, a finite velocity spread will again cause a reduction of the fringe visibility.
 
\paragraph{Optimized grating roll:}
 Since both the gravitational phase shift and the Coriolis phase shift are velocity dependent, and since the sign of the gravitational phase can be chosen by the sign of the common roll angle, one can find an angle where both effects cancel exactly for one specific velocity and are still largely compensated in a finite velocity band around that velocity \cite{Fein2020a}. The optimal correction angle is determined by the geographic latitude of our lab: 
\begin{equation}
\theta_0=-\Omega_E \sin(\theta_L) v/g
\end{equation}
Given that $\Omega_E=72$ µrad/s, $g=9.81$ m/s$^2$ and $\theta_L=$48° for Vienna, a cluster beam of $v=100$ m/s will require a common grating roll angle of
$\theta_0=-0.55$\,mrad which can be set easily and accurately.   

\begin{figure}[h]
    \centering
    \includegraphics[width=0.47\textwidth]{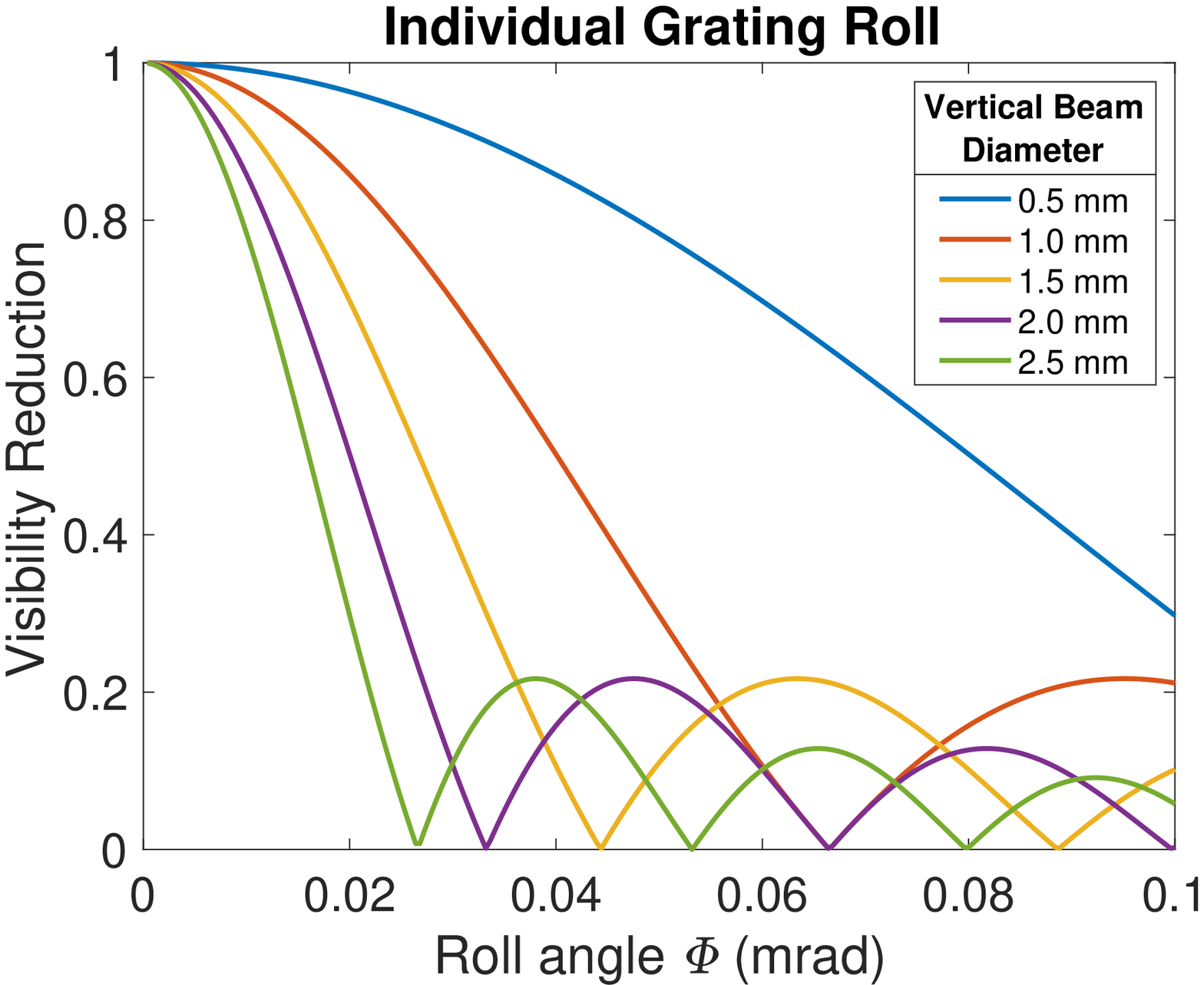}
        \includegraphics[width=0.5\textwidth]{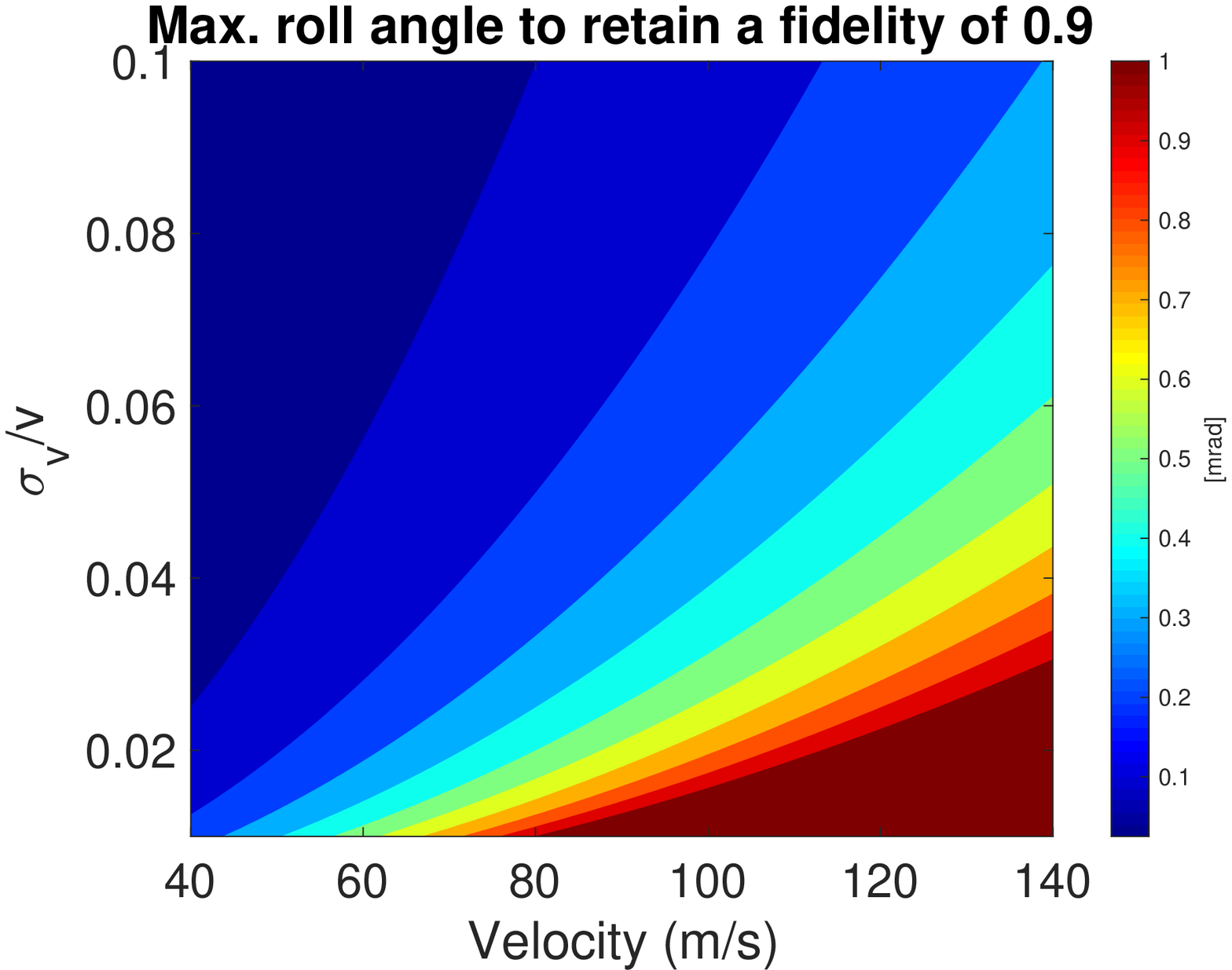}
    \caption{Left: Visibility reduction as a function of grating roll misalignment with respect to each other for different vertical beam diameters. Right: Maximum permissible roll misalignment of the gratings with respect to gravity versus particle velocity and velocity spread.}
    \label{fig:my_label}
\end{figure}

\paragraph{Independent grating vibrations:}
A lateral position shift of any of the gratings enters the total interferometer phase shift like
\begin{equation}
    \Delta \varphi =k_d (x_1-2x_2+x_3) = k_d (A_1 sin(\omega_1 t)-2 A_2 sin(\omega_2 t)+A_3 sin(\omega_3 t)).
\end{equation}
Random x-oscillations of any of these gratings with a fixed amplitude $A_i$ but white noise frequency spectrum and random phase then reduces the total fringe visibility by a factor
\begin{equation}
R_{\mathrm{Vib}}=\left|J_0\left(\frac{2 \pi}{d} A_{1,3}\right)\right|^2 \cdot\left|J_0\left(\frac{4 \pi}{d} A_2\right)\right|,
\end{equation}
where $J_0$ is the zeroth order Bessel function.

\paragraph{Common mode vibrations:}
In order to suppress excitations of the individual gratings, the entire interferometer is mounted on a massive Invar bar which is suspended by piano wires, isolated by a stack of plates and Teflon balls and damped by eddy current breaks. 
This still allows for common mode motion of amplitude $A$ and frequency $\omega$ which would reduce the fringe visibility by 
\begin{equation}
R_{\mathrm{vib}}=\left|J_0\left(\frac{8 \pi A}{d} \sin ^2\left(\frac{\omega L}{2 v_z}\right)\right)\right|.
\end{equation}

\begin{figure}[h]
    \centering
    \includegraphics[width=0.49\textwidth]{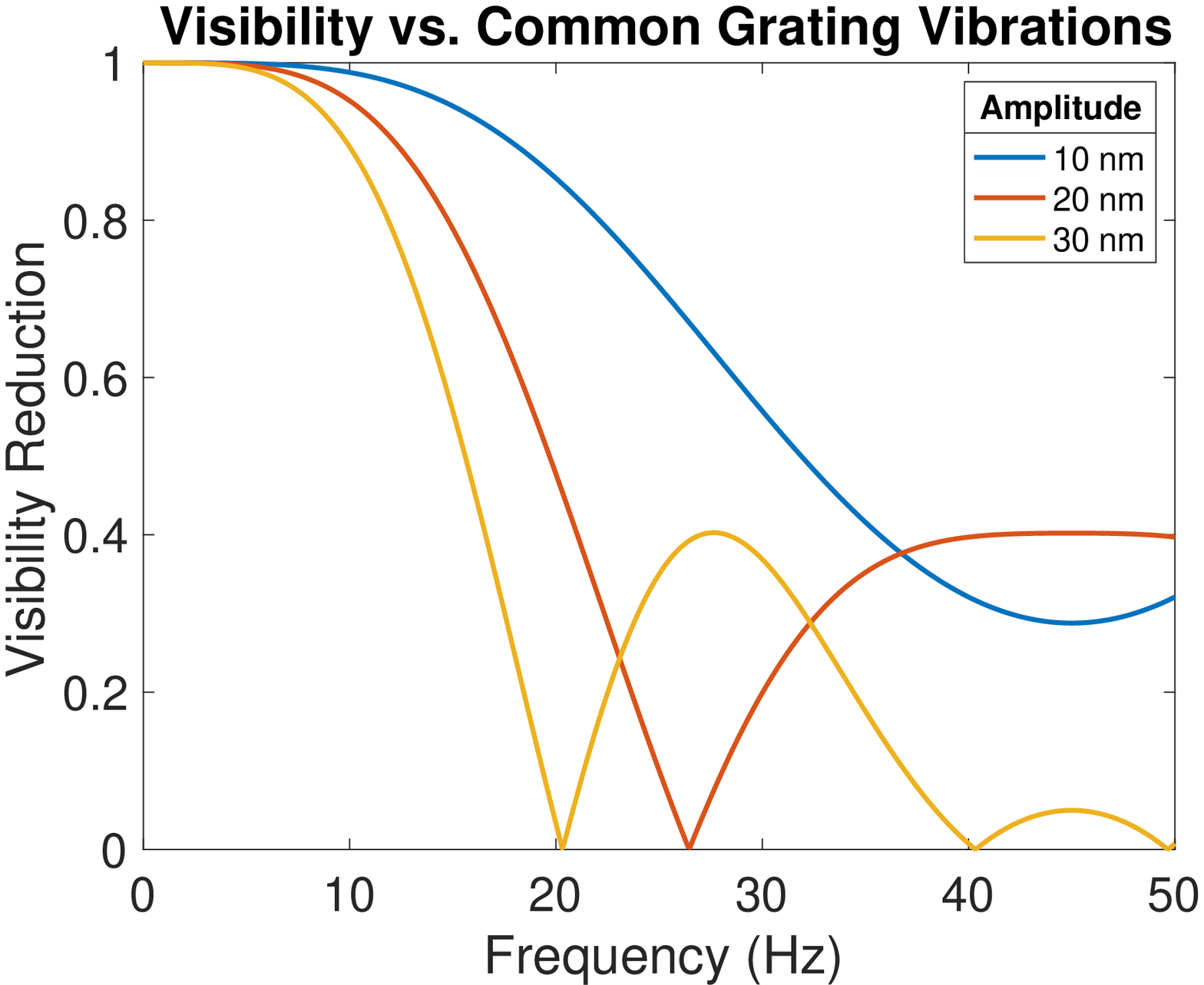}
        \includegraphics[width=0.49\textwidth]{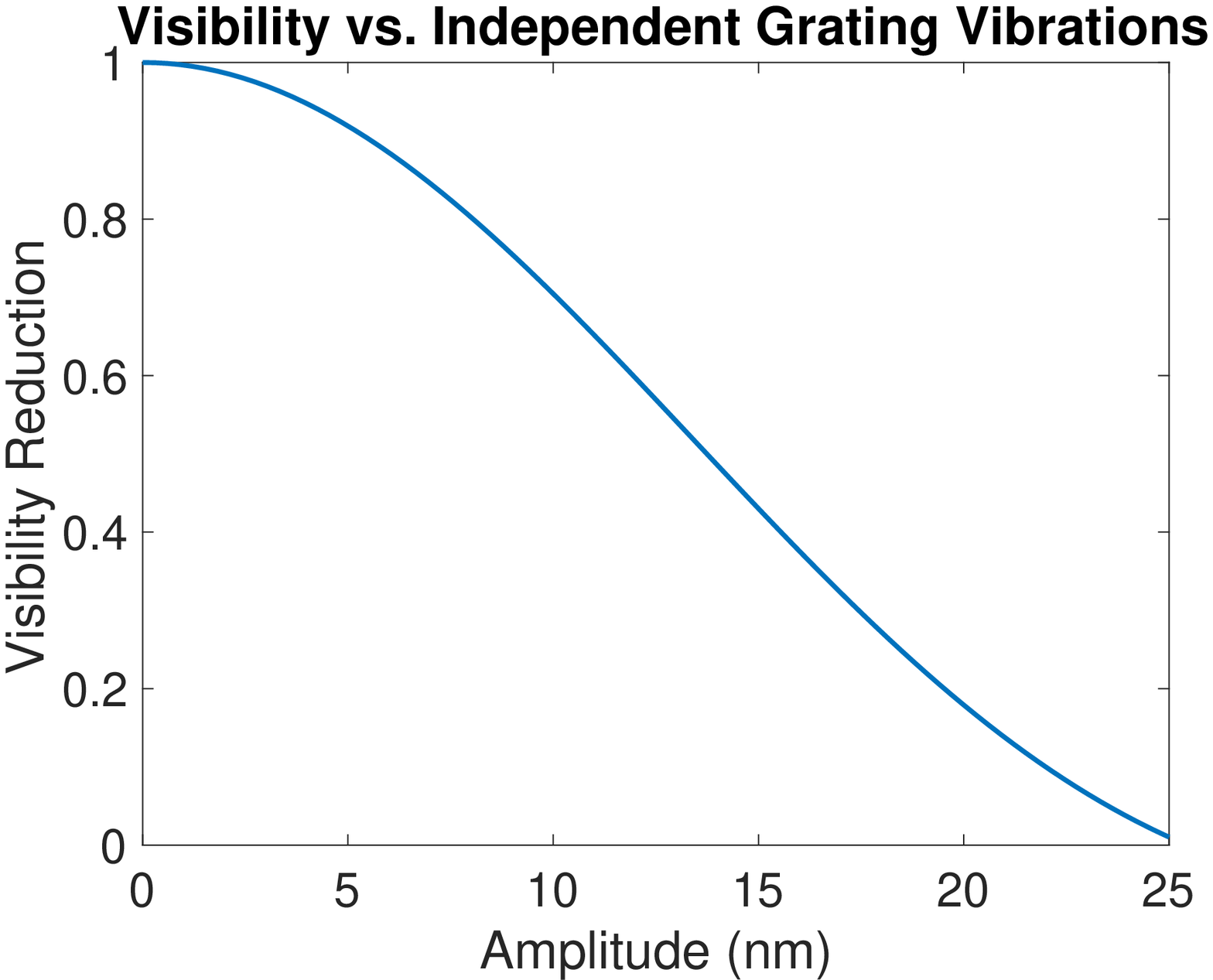}
    \caption{Left: Visibility reduction as a function of frequency for different amplitudes of common grating vibrations. A monochromatic velocity of 90 m/s is assumed here. At certain frequencies, visibility revivals can be observed. Right: Visibility reduction as a function of amplitude for independent grating vibrations. For individual grating vibrations, the visibility reduction is independent of the vibration frequency.}
    \label{fig:common_vib}
\end{figure}


\section{CONCLUSION}
Matter-wave interferometry with metal clusters is a promising path towards quantum experiments with objects of high mass, whose separation can be on the scale of 100\,nm, i.e. substantially larger than the object itself. The alignment criteria outlined in this work demonstrate that, while challenging, it is feasible to construct an interferometer capable of studying particles with masses of 100 kDa using state-of-the-art technologies. By collimating the molecular beam, particularly the beam width $b$ in the transverse direction, the alignment criteria can be met. A magnetron sputtering source can already provide sufficient flux to work with collimated beams and metal clusters up to 1 MDa. While increasing the degree of collimation will make the alignment criteria less stringent, it will also decrease the signal-to-noise ratio. In practice the collimation will be a trade-off between the flux and alignment feasibility.
Furthermore, this study highlights the importance of ongoing advancements in molecular beam source and detection techniques in order to achieve the high and stable molecular flux required to meet the more stringent alignment requirements necessary to study clusters with masses up to 1 MDa and beyond. 
The reward will be new interferometric exclusions of non-linear alternatives to quantum mechanics, as well as a powerful tool that can serve material science with refined measurements of properties of isolated nanomaterials.     

\begin{table}[h]
\centering
\begin{tabular}{c|c|c}
Degree of freedom        & Restriction & Alignment requirement \\
\hline
Grating period           &    $N < \frac{d}{10\Delta d}$        &      $N \leq 5 \times 10^6 d$     \\

Grating separation       &  $\frac{\Delta L}{L}< \frac{1}{N}$         &   $\Delta L<133$ µm                \\

Grating roll             &    $R_\vartheta<0.9$         &      $\Delta\vartheta <0.02$ mrad                \\

Grating yaw              &         $10\cdot\Delta\Phi <\frac{d}{4w_x}$  &         $\Delta\Phi<$0.2 mrad             \\

Grating pitch            &    $  \Delta \Theta\cdot w_y <\Delta L $       & $\Delta\Theta<100$ mrad                     \\        
Wave front shape         &       $\frac{w(z)^2}{R(z)}\ll d$      &    $z<2.7$ mm            
\end{tabular}
\vspace{0.3cm}
\caption{Summarized alignment criteria for an all-optical Talbot-Lau inteferometer with a grating period of $d=133$ nm separated by $L=1$ m, a molecular beam height and width of $H=1$ mm and $b=1$ mm and a laser beam waist of $w_x=15$ µm and $w_y=1.5$ mm required to obeserve quantum interference of yttrium clusters with a mass of $m=100$ kDa and a forward velocity of $v_x=100$ m/s. }
\end{table}
 
\acknowledgments 
We acknowledge funding by the Austrian Science Funds in FWF project No. P 32543-N  (MUSCLE) as well as support by the Gordon and Betty Moore Foundation within the project No. 10771 (ELUQUINT).  

\bibliography{main.bib} 
\bibliographystyle{spiebib} 

\end{document}